\documentclass[a4paper]{article}
\usepackage[utf8]{inputenc}
\usepackage{url}
\usepackage{csquotes}

\title{Wikipedia's Balancing Act: A Tool for Collective Intelligence or Mass Surveillance?}
\author{\textbf{Simon Liu}
        \\
        \footnotesize
        BCom, LLB
        \\
        \footnotesize
        The University of Sydney
        }
\date{12 December 2022}

\begin{document}

\maketitle

\begin{abstract}
\noindent Wikipedia has evolved beyond its original function as an online encyclopedia in an increasingly complex data-driven society. The social platform is met with a balancing act between collective intelligence and mass surveillance; processes need to be developed to protect individuals and the community from government mass surveillance without sacrificing the important contributions made through prohibited anonymous communication software. Case studies are provided from NSA government surveillance practices, the anti-SOPA legislation movement, and research that covers Wikipedia’s involvement with participatory journalism, disinformation, self-censorship, and the use of Tor. This paper proposes that a common ground can be developed between individuals, public and private institutions through future research in socio-cultural anthropology and policy frameworks around data retention and government accountability. Wikipedia is used as an example within the US intelligence community as a complex organisation that can adapt to changes through its iterative nature, which draws insight into how policy frameworks can be future-proofed. Finally, this paper is a wake-up call to individuals, private institutions, and governments to remain vigilant about the storage and use of personal information as a result of contributing to online communities.
\end{abstract}

\section{Introduction}
Educational websites that exist today not only function as sources of information but also as social platforms. In particular, Wikipedia’s unique open-editing model invites anyone to contribute information to Wikipedia pages and articles regardless of their expertise, creating a community-generated social space similar to Facebook, Instagram, Reddit, and Twitter (Di Lauro \& Johinke, 2017). Founded in 2001, Wikipedia as an online encyclopedia has played an integral role in disrupting the landscape of traditional social, political, legal, and economic institutions through its collective intelligence, influencing the anti-SOPA legislation movement in 2012 and engaging in participatory journalism (Greenstein \& Zhu, 2012; Pasquale, 2015; Konieczny, 2014; Lih, 2004). However, with great power comes great responsibility. As one of the most visited and widely cited websites in the world with over 1.9 edits made per second, Wikipedia’s high search engine ranking has the potential to amplify disinformation, prevalent in ongoing news events like natural disasters (Wikipedia, 2020; Rosenzweig, 2006; Sáez-Trumper, 2019). In fact, studies have shown that humans are worse when compared to bots at distinguishing between legitimate and hoax articles (Kumar, West, \& Leskovec, 2016). The impact of collective intelligence on society will be examined, presented from the perspectives of all types of stakeholders. \vskip 0.2in

\noindent In an increasingly data-driven society, the question of organisational trust on our personal data needs to be explored: what kind of personal information is being collected? In the context of Wikipedia, bots are used as editorial oversight and control, contributing about 20\% of all edits (Geiger, 2017). These data-driven algorithms that parse through information exists across all organisations, including government intelligence communities as exposed by NSA whistleblower Edward Snowden, who leaked government surveillance practices such as the involvement of technology firms with PRISM, the mass data collection system which expressly identified Wikipedia as a target of surveillance (Wales \& Tretikov, 2015). Public awareness of increased mass surveillance amounted to a ‘chilling effects’ situation — “that certain state acts may chill or deter people from exercising their freedoms or engaging in legal [online] activities” (Penney, 2016). Penney’s chilling effects studies on Wikipedia contributors of controversial or sensitive topics explain a sharp decline in the number of edits of terrorism-related articles post-Snowden leaks, which is attributable to: the uncertainty surrounding the vague and ambiguous application of the law; or alternatively, to avoid being labelled or tracked by the government as non-conformists or criminals (Schauer, 1978; Solove, 2006). Thus, the use of Tor, an anonymous communication software, will be explored to understand the motivation behind Wikipedia contributors masking their digital footprint and whether Tor users are more likely to edit articles considered to be controversial or sensitive (Forte, Andalibi, \& Greenstadt, 2017; Tran, Champion, Forte, Hill, \& Greenstadt, 2020). \vskip 0.2in

\noindent Finally, what does the future hold for Wikipedia? How do we resolve the tension between collective intelligence and mass surveillance? The Oxford Dictionary of English definition of “balancing act” (Stevenson, 2010, p. 123) encapsulates this idea well — Wikipedia needs to perform an action or activity that requires a delicate balance between different situations or requirements. Future implications will be discussed with possible solutions, in addition to a case study by Andrus (2005) on how Wikipedia’s iterative process, if replicated within the intelligence community, can overcome the increasingly complex and non-linear national security environment.

\section{Collective Intelligence}

\subsection{Why Contribute?}
There are many reasons why individuals contribute to online communities, which include both intrinsic and extrinsic motivation. From generalised reciprocity to social recognition, users will continue to contribute and learn from others within a social context (Tedjamulia, Dean, Olsen, \& Albrecht, 2005). There are two factors hypothesised to increase the level of participation: having a personal goal and commitment to the project. For example, an individual is more inclined to share their knowledge on a platform that encourages freedom of speech.

\subsection{Linus's Law}
Wikipedia’s open-source collaborative platform is a successful example of collective intelligence in the online community. Notably, Wikipedia articles have been cited in multiple court cases and judicial opinions around the world given the platform's ability to provide up-to-date crowdsourced information that can be readily relied upon as supplementary material (Wikipedia, 2020; Rukundo, 2019). Previous citations include geographic information and definitions of phrases (Miller \& Murray, 2010). Linus’s Law explains the process of collective intelligence; having a large enough number of contributors will eventually replace the existing inaccurate information with input that is verifiable and higher in quality (Raymond, 1999; Greenstein \& Zhu, 2012). However, there are practical limitations in this concept. The attention and quality assurance of each article is not evenly distributed. In addition, studies have shown that there are inherent cultural and gender bias on Wikipedia (Hube, 2017). Researchers have observed an increase in fake news coverage during the 2016 United States presidential campaign in social communities such as Facebook and Twitter, resulting in ‘filter bubbles’ and ‘echo chambers’ filled with misinformation that reinforces an individual's preconceived notion (Guess, Nyhan, \& Reifler, 2018). In particular, Wikipedia functions like an amplifier through search engines such as Google, which captures a snapshot of a Wikipedia article that may be inaccurate at the time, and the inaccuracies are broadcasted instantly to internet users worldwide (Rosenzweig, 2006). Another implication of Wikipedia is its effect on education with its tendency to appear on multiple front page search results. 87.5\% of Australian students are reliant on Wikipedia as an information source, however, survey data of the students has demonstrated its use as supplementary material, rather than as a primary source (Selwyn \& Gorard, 2016).

\subsection{Neutral Point of View}
Wikipedia’s neutral point of view policy plays a significant role in preventing disinformation online. As a defining principle of Wikipedia, it is a guideline that all contributors should follow in good faith — to write fairly and reliably to provide a complete representation of the topic, and without editorial bias, presenting all points of views proportionally (Cap, 2012). Wikipedia’s neutral point of view policy stands to the test of time since the creation of the platform, and has continued to function as a cornerstone of the Wikipedia ideology. It is understood that the barriers around editorial bias are “not firmly structural, but rather socially constructed”, which demonstrates that a reduction in disinformation on Wikipedia over time can be attributed to an increasing emphasis on community participation to moderate and prevent disinformation on online platforms (Livingstone, 2010).

\subsection{Disinformation}
Despite Wikipedia's role in tackling disinformation, it is still necessary to consider how disinformation could be easily spread through Wikipedia. For example, ongoing news events like natural disasters can be exploited due to the lack of up-to-date information and therefore, disinformation can appear across the search results of search engines due to its high page ranking (Sáez-Trumper, 2019). This led to a spread of disinformation as seen in Croatia Wikipedia, where certain extremist or fascist events have been represented more positively than in its English counterpart (Vrandečić, 2019). Human experiments have also demonstrated the poor ability to distinguish between legitimate and hoax Wikipedia articles with a 66\% detection rate, attributable to human bias of perceiving well-formatted articles as legitimate. An example is the “Bicholim conflict”, an article that achieved a Good Article status for half a decade despite being a hoax article (Kumar et al., 2016).

\subsection{Anti-SOPA Legislation Movement}
The rise of user-generated content websites has resulted in participatory journalism, operating similar to traditional media companies in the process of gathering, filtering, and visualising content to be presented in a digestible format (Lih, 2004). Wikipedia’s unique open-editing model is faster and more efficient than traditional media companies, for example, the unbureaucratic process between reading ongoing news events and editing the article provides live updates to readers worldwide (Stein, 2013). These websites have also participated in online activism, which held the power to influence international policy such as the anti-SOPA legislation movement in 2012 (Konieczny, 2014). The “Stop Online Piracy Act” Bill 2011 (US) was a controversial legislation, if enacted, could impose liability on online communities for the actions of its users if alleged of any online copyright infringement, bringing about a chilling effect of site-wide censorship that can be easily exploited on a publicly editable platform such as Wikipedia. Wikipedia contributors played a role in gaining international protest movement; a majority vote on the Wikipedia:SOPA poll supported Wikipedia’s 24-hour page blackout joined by Google and other websites, and shortly afterwards, the bill was postponed indefinitely (Wikipedia, 2020). A survey concluded that Wikipedia contributors are motivated by ‘free speech’: a desire to collectively contribute, derived from a free-information ideology of democratic values (Nov, 2007). Yet, the guidelines of Wikipedia is paradoxical. The proud platform that encourages open discussion and collaboration amongst its users discourages the use of the term ‘terrorist’, which was also adopted by news organisation Reuters in an apology resulting from the aftermath of the September 11 attacks (Wikipedia, 2020; Glocer \& Linnebank, 2001). This demonstrates Wikipedia’s evolution of what it was originally as an online encyclopedia, into a social, political, legal, and economic conglomerate.

\subsection{Future of Collective Intelligence}
It has been demonstrated that an unregulated ‘free market’ of contributions led to a plethora of high quality legitimate and hoax articles. Wikipedia’s own guidelines are unable to provide oversight to problems that have been originally out of the scope of its original function as an online encyclopedia. If Wikipedia is unable to cope with the demands and consequences of its contributions, then where does the future of collective intelligence lead? Faucher, Everett, and Lawson’s (2008) definition of complexity theory which addresses this question will be explored in the fourth section of this paper, that asks whether online communities are able to adapt rapidly to unpredictable future complex environments. Although Wikipedia does not officially recognise itself as a social networking site, future-proof policies must be developed around disinformation and its role as a platform. But before we look at the future implications and possible solutions, Wikipedia has another force to balance other than collective intelligence.

\section{Mass Surveillance}

\subsection{Within the \enquote{Black Box}}
Data-driven algorithms exists across all aspects of our lives, from applying for a credit card to using voice assistants on our devices. They play an important but pervasive role in organising, ordering, governing, and gatekeeping our personal information inside the ‘black box’ which hides the decision-making source codes that have an extensive impact on society with minimal accountability (Geiger, 2017; Pasquale, 2015). Currently, researchers are focused on the question of the future of technologies in transitioning to open-source software (Chesbrough, 2017; Coelho \& Valente, 2017). Literature papers have overlooked using Wikipedia as a case study which provides researchers with current insight on the issues surrounding data-driven algorithms that exist beyond the black box. Either way, our trust in organisations on our personal data is inevitable and instead, we should ask ourselves: what kind of personal information is being collected about me and what purpose is it being used for? Empirical evidence suggests that websites are increasingly intrusive, personalising our view of the internet and ultimately, our perspectives in a capitalistic privacy regulation environment (Cecere \& Rochelandet, 2013; Hui \& Png, 2006).

\subsection{Chilling Effects}
This leads to the discussion of how a ‘chilling effects’ situation undermines the collective effort of individuals. Snowden’s leaks from 2013 to 2014 of NSA ‘upstream’ surveillance system was the catalyst for public awareness and a change in public perceptions around trust and accountability with the government (Szoldra, 2016). Shortly afterwards, Wikimedia Foundation and other organisations filed a lawsuit against the NSA and other named state actors alleging mass surveillance of Wikipedia users, which violated the fundamental rights of privacy, freedoms of expression and association. Jimmy Wales and Lila Tretikov, the co-founder of Wikipedia and former executive director of Wikimedia Foundation respectively, published a joint co-ed contribution on The New York Times highlighting the NSA’s ability to parse through a user’s contribution to Wikipedia, thus tracking their digital footprint and creating a profile based on their personal information (Wales \& Tretikov, 2015). Therefore, it would be reasonable for contributors to be less willingly to share their information out of fear of persecution, resulting in a domino effect that disrupts the collective intelligence cycle. \vskip 0.2in

\noindent Empirical findings support the chilling effects theory on Wikipedia contributors post-Snowden leaks; there was a surveillance-related chill that caused the general trend reversal and sudden drop in the number of edits of terrorism-related articles (Penney, 2016). The findings also support the opposite; other topics on Wikipedia were not impacted by the chill as it would not have been likely to raise privacy concerns. However, there are limitations as noted by Penney; the research was conducted within a two-year period which is not determinant of a long-term chilling effects situation. Moreover, the study was based on English Wikipedia, to which the evidence of chilling effects may not be applicable in other Wikipedia languages. A further study conducted by Büchi et al. (2019) has suggested that a subtle and long-term chilling effect may exist from organisations profiling our information. Other studies have suggested that a consistent chilling effect is not wide-spread across different languages, but a wider time-period needs to be considered to measure its long-term impact (Xie, Johnson, \& Gomez, 2019).

\subsection{Wikipedia and Tor}
Open-collaboration projects are trying to address the problem of maintaining user participation over time, hoping to encourage a diverse range of contributors to ensure equal representation, quality, and coverage (Forte et al., 2017; Sydow, Baraniak, \& Teisseyre, 2017). Since the public is more aware about their vulnerabilities when using the internet post-Snowden leaks, Wikipedia contributors realise that their actions are publicly logged and made available through public databases (Wikimedia Foundation, 2020). Furthermore, Wikipedia has blocked the use of Tor, an anonymous communication software, when editing articles to prevent abuse and vandalism by individuals that are non-identifiable (Wikipedia, 2020). The Tor IP address blocks were manually performed by administrators, but soon after, it was replaced with bots which led to slippages in the system (Tran et al., 2020). Analysis of the Tor IP addresses have demonstrated that topics considered more controversial and sensitive (i.e. politics, technology, and religion) were more likely to be edited by Tor users, and at a higher quality, highlighting the importance and value of anonymous contributions that have been largely neglected. Additional research has found that one-time contributions of anonymous users provided the highest quality contributions out of all users, who tend to be selective with the topics that they wrote (Lichtenstein \& Parker, 2009; Huang, Zhu, Du, \& Lee, 2016). Similar to the limitation of the chilling effect studies on Wikipedia, the studies were conducted on English Wikipedia, to which the results may not extend across other Wikipedia languages. \vskip 0.2in

\noindent Without getting into the technicalities of proposed machine learning techniques, onion routing systems such as Tor do not guarantee complete anonymity (Barker, Hannay, \& Bolan, 2010; Mercaldo \& Martinelli, 2017). In order to propose a solution around anonymous editing, it is important to consider the motivations behind self-censorship. Interviews were conducted to Wikipedia contributors that did edit and did not edit through Tor; the findings highlighted that the main concern of contributors was the loss of privacy and surveillance from governments (Forte et al., 2017). In particular, contributors that had taken on leadership roles (as time passed by) were more concerned about their ability to handle privacy threats throughout life changes such as starting a family or applying for a job or school.

\subsection{Future of Mass Surveillance}
Loss of privacy and mass surveillance are concerns that exist in online communities. Processes need to be developed to protect individuals and the community from future chilling effects without sacrificing the important contributions made through prohibited anonymous communication software. In an ever-changing social, political, legal, and economic environment, internet contributors must remain vigilant of their actions, which may have future negative consequences.

\section{Future Implications and Possible Solutions}

\subsection{Further Research To Be Conducted}
Many studies cited in this paper have not demonstrated its application in other Wikipedia languages (Hube, 2017; Penney, 2016, Lichtenstein \& Parker, 2009; Huang et al., 2016). A better understanding of whether chilling effects occurs in other languages serves as guidance to combat the ‘thawing’ of collective intelligence. For example, if the chilling effects situation cannot be observed in other languages such as Chinese and Japanese as a result of their collectivist culture, then possible solutions may involve developing a stronger online community that reinforces notions of group ideology. Likewise, the number of tor-based edits observed in other languages may provide a better understanding of which cultures are more susceptible to mass surveillance. We propose that future research in socio-cultural anthropology may serve as guidance to balance Wikipedia’s need to maintain its collective intelligence without deterring individuals from anonymously contributing as a result of government mass surveillance practices (Gloor et al, 2017).

\subsection{Complexity Theory}
A paper published by Andrus (2005) in the Central Intelligence Agencies’ journal “Studies in Intelligence” draws insight into the complementary nature of collective intelligence and mass surveillance, and how it can be applied in a future setting with an example provided within government intelligence communities. The intelligence community operates in a real-time strategic environment; the total duration of the intelligence - decision - implementation cycle "can be as short as 15 minutes” (Andrus, 2005). Otherwise known as complexity theory, in an increasingly complex and non-linear national security environment, intelligence processes must implement iterative methodologies to incorporate continuous learning and adaptation. Wikipedia as a complex adaptive organisation is used as a case study to highlight a flexible and loose management structure in collaboration with complex human-computer interactions (Faucher et al., 2008). Here, Wikipedia contributors are granted more autonomy compared to traditional organisations, in which the cumulative efforts by the community has resulted more efficient and increased information-sharing, whereby talk pages also serve as a means to discuss and share different perspectives with other collaborators. To replicate a complex adaptive system such as Wikipedia, the following question should be considered: how can a community introduce policies that modify its environment to adapt to rapid and unpredictable changes? Andrus outlines the six key elements of a complex adaptive system: self-organisation, emergence, relationships, feedback, adaptability, and non-linearity (Turnbull \& Guthrie, 2019). Applying these concepts to the intelligence community, Andrus theorises that intelligence officers are able to act more independently and be knowledgeable through information-sharing, receive more direct feedback throughout the intelligence - decision - implementation cycle process, and finally, be made more aware of the bigger picture of the organisation, reinforcing group ideology. \vskip 0.2in

\noindent Later papers have discussed Intellipedia, a Wiki platform introduced in 2006 that is used internally within the Central Intelligence Agency (Werbin, 2011). Intellipedia has become an integral part of the intelligence community, improving collaboration and technological competence of intelligence officers from their bottom-up participation. Real-life successes in North Korea, Iraq, and Nigeria were recognised in Hickman’s (2010) book. As necessary, limitations are to be discussed. Further research should take into consideration of whether dynamic intelligence should replace bureaucratic intelligence processes (Chomik, 2012). In a broader context, is the future of complex collective intelligence processes an open-source environment?

\subsection{The \enquote{Balancing Act}}
The paradigm of the balance between collective intelligence and mass surveillance is left unexplored, especially Wikipedia’s role in the process. Wikipedia will continue to grow socially, politically, legally, and economically. Likewise, society will observe an increase in self-censorship from government mass surveillance practices. Towards a better future involves understanding the complementary nature of the two forces, to develop a policy framework that imposes stricter data retention laws on users that wish to remain anonymous, but places the onus on the government to demonstrate its warranted access to user information (Khatri \& Brown, 2010).

\section{Conclusion and Final Remarks}
Wikipedia is introduced in this paper to outline its current and potential future practices around pertinent issues in today’s society. The two issues introduced are collective intelligence and mass surveillance. This paper draws examples from Wikipedia’s involvement with individuals, private institutions, and governments in order to discuss about how Wikipedia has balanced the two opposing forces and propose methods to find common ground between the different stakeholders through policy-making. But as always, there is no safe harbour. Complex changes are likely to occur in every organisation, that will have long-lasing detriment internally and externally if not dealt with pre-emptively. \vskip 0.2in

\noindent The reader might perceive the tone of this paper as negative. It is essential to be pragmatic about the way technologies can be utilised for the benefit of public and private institutions; it is inevitable that we are increasingly dependent on the conveniences provided as such. Overall, we should remain optimistic as a society about what the future of technology holds. Internal policies of online communities and government legislation will be strictly regulated in supporting the transition to a more transparent environment with accountability from all types of stakeholders. \vskip 0.2in

\noindent Finally, the best time to act is now. This paper serves as a wake-up call to individuals, private institutions, and governments to remain vigilant about the storage and use of personal information as a result of contributing to online communities. A balancing act is to be performed.

\section*{Acknowledgements}
I would like to thank my supervisor, Dr. Frances Di Lauro, for providing me with the opportunity to partake in this research topic.

\section*{References}
Andrus, D. C. (2005). The Wiki and the Blog: Toward a Complex Adaptive Intelligence Community. \textit{Studies in Intelligence}, 49(3). \vskip 0.1in

\noindent Barker, J., Hannay, P., \& Bolan, C. (2010). Using Traffic Analysis to Identify Tor Usage — A Proposed Study. In \textit{Security and Management} (pp. 620-623). \vskip 0.1in

\noindent Büchi, M., Fosch V. E., Lutz, C., Tamò-Larrieux, A., Velidi, S., \& Viljoen, S. (2019). Chilling Effects of Profiling Activities: Mapping the Issues. \textit{Available at SSRN 3379275}. \vskip 0.1in

\noindent Cap, C. H. (2012). Towards Content Neutrality in Wiki Systems. \textit{Future Internet}, \textit{4}(4), 1086-1104. \vskip 0.1in

\noindent Cecere, G., \& Rochelandet, F. (2013). Privacy intrusiveness and web audiences: Empirical evidence. \textit{Telecommunications Policy}, \textit{37}(10), 1004-1014. \vskip 0.1in

\noindent Chesbrough, H. (2017). The Future of Open Innovation: The future of open innovation is more extensive, more collaborative, and more engaged with a wider variety of participants. \textit{Research-Technology Management}, \textit{60}(1), 35-38. \vskip 0.1in

\noindent Chomik, A. (2012). Spies Wearing Purple Hats: \textit{The use of social computing to improve information sharing inside the Intelligence Community of the United States} (Master's Thesis, University of Calgary). \vskip 0.1in

\noindent Coelho, J., \& Valente, M. T. (2017, August). Why Modern Open Source Projects Fail. In \textit{Proceedings of the 2017 11th Joint Meeting on Foundations of Software Engineering} (pp. 186-196). \vskip 0.1in

\noindent Di Lauro, F., \& Johinke, R. (2017). Employing Wikipedia for good not evil: innovative approaches to collaborative writing assessment. \textit{Assessment \& Evaluation in Higher Education}, \textit{42}(3), 478-491. \vskip 0.1in

\noindent Faucher, J. B. P., Everett, A. M., \& Lawson, R. (2008). A Complex Adaptive Organization Under the Lens of the LIFE Model: The Case of Wikipedia. In \textit{Fourth Organizational Studies Summer Workshop: “Embracing Complexity: Advancing Ecological Understanding in Organization Studies”}. \vskip 0.1in

\noindent Forte, A., Andalibi, N., \& Greenstadt, R. (2017, February). Privacy, Anonymity, and Perceived Risk in Open Collaboration: A study of Tor Users and Wikipedians. In \textit{Proceedings of the 2017 ACM Conference on Computer Supported Cooperative Work and Social Computing} (pp. 1800-1811). Association for Computing Machinery. \vskip 0.1in

\noindent Geiger, R. S. (2017). Beyond opening up the black box: Investigating the role of algorithmic systems in Wikipedian organizational culture. \textit{Big Data \& Society}, \textit{4}(2), 2053951717730735. \vskip 0.1in

\noindent Glocer, T., \& Linnebank, G. (2001, October 2). Letter to the editors of certain US newspapers. \textit{Reuters}. Retrieved from \url{https://web.archive.org/web/20011002123400/http://about.reuters.com/statement3.asp} \vskip 0.1in

\noindent Gloor, P. A., Marcos, J., De Boer, P. M., Fuehres, H., Lo, W., \& Nemoto, K. (2017). Cultural Anthropology through the Lens of \textit{Wikipedia}. Social Network Analysis: Interdisciplinary Approaches and Case Studies, 245. \vskip 0.1in

\noindent Greenstein, S., \& Zhu, F. (2012). \textit{Collective Intelligence and Neutral Point of View: The Case of Wikipedia} (No. w18167). National Bureau of Economic Research. \vskip 0.1in

\noindent Guess, A., Nyhan, B., \& Reifler, J. (2018). Selective Exposure to Misinformation: Evidence from the consumption of fake news during the 2016 U.S. presidential campaign. \textit{European Research Council}, \textit{9}(3), 4. \vskip 0.1in

\noindent Hickman, G. R. (2010). \textit{Leading Change in Multiple Contexts: Concepts and Practices in Organizational, Community, Political, Social, and Global Change Settings}. Sage Publications. \vskip 0.1in

\noindent Huang, L., Zhu, Q., Du, J. T., \& Lee, B. (2016). Exploring the dynamic contribution behavior of editors in wikis based on time series analysis. \textit{Program}, \textit{50}(1), 41-57. \vskip 0.1in

\noindent Hube, C. (2017, April). Bias in Wikipedia. In \textit{Proceedings of the 26th International Conference on World Wide Web Companion} (pp. 717-721). \vskip 0.1in

\noindent Hui, K. L., \& Png, I. P. L. (2006). The Economics of Privacy. In \textit{Economics and Information Systems, Volume 1 (Handbooks in Information Systems)} (Chapter 9). \vskip 0.1in

\noindent Khatri, V., \& Brown, C. V. (2010). Designing Data Governance. \textit{Communications of the ACM}, \textit{53}(1), 148-152. \vskip 0.1in

\noindent Konieczny, P. (2014). The day Wikipedia stood still: Wikipedia’s editors’ participation in the 2012 anti-SOPA protests as a case study of online organization empowering international and national political opportunity structures. \textit{Current Sociology}, \textit{62}(7), 994-1016. \vskip 0.1in

\noindent Kumar, S., West, R., \& Leskovec, J. (2016, April). Disinformation on the Web: Impact, Characteristics, and Detection of Wikipedia Hoaxes. In \textit{Proceedings of the 25th international conference on World Wide Web} (pp. 591-602). International World Wide Web Conferences Steering Committee. \vskip 0.1in

\noindent Lichtenstein, S., \& Parker, C. M. (2009). Wikipedia model for collective intelligence: a review of information quality. \textit{International Journal of Knowledge and Learning}, \textit{5}(3-4), 254-272. \vskip 0.1in

\noindent Lih, A. (2004). Wikipedia as Participatory Journalism: Reliable Sources? Metrics for evaluating collaborative media as a news resource. \textit{Nature}, \textit{3}(1), 1-31. \vskip 0.1in

\noindent Livingstone, R. M. (2010). Let’s Leave the Bias to the Mainstream Media: A Wikipedia Community Fighting for Information Neutrality. \textit{M/C Journal}, \textit{13}(6). \vskip 0.1in

\noindent Mercaldo, F., \& Martinelli, F. (2017, September). Tor Traffic Analysis and Identification. In \textit{2017 AEIT International Annual Conference} (pp. 1-6). IEEE. \vskip 0.1in

\noindent Miller, J. C., \& Murray, H. B. (2010). Wikipedia in Court: When and How Citing Wikipedia and Other Consensus Websites Is Appropriate. \textit{St. John's Law Review}, \textit{84}(2), 633. \vskip 0.1in

\noindent Nov, O. (2007). What Motivates Wikipedians?.\textit{ Communications of the ACM}, \textit{50}(11), 60-64. \vskip 0.1in

\noindent Pasquale, F. (2015). The Black Box Society: The Secret Algorithms That Control Money and Information. Cambridge, Massachusetts; London, England: Harvard University Press. \vskip 0.1in

\noindent Penney, J. W. (2016). Chilling Effects: Online Surveillance and Wikipedia Use. \textit{Berkeley Technology Law Journal}, \textit{31}(1), 117-182. \vskip 0.1in

\noindent Privacy policy. (n.d.). In Wikimedia Foundation. Retrieved August 19, 2020, from \url{https://foundation.wikimedia.org/wiki/Privacy_policy} \vskip 0.1in

\noindent Raymond, E. (1999). The Cathedral and the Bazaar. \textit{Knowledge, Technology \& Policy}, \textit{12}(3), 23-49. \vskip 0.1in

\noindent Rosenzweig, R. (2006). Can History Be Open Source? \textit{Wikipedia} and the Future of the Past. \textit{The Journal of American History}, \textit{93}(1), 117-146. \vskip 0.1in

\noindent Rukundo, S. (2019). Wikipedia in the Courts: An examination of the citation of Wikipedia in judicial opinions in Uganda. \textit{Computer Law \& Security Review}, \textit{35}(5), 105316. \vskip 0.1in

\noindent Sáez-Trumper, D. (2019). Online Disinformation and the Role of Wikipedia. \textit{ArXiv}, \textit{abs/1910.12596}. \vskip 0.1in

\noindent Schauer, F. (1978). Fear, Risk and the First Amendment: Unraveling the Chilling Effect. \textit{Boston University Law Review}, \textit{58}(5), 685-732. \vskip 0.1in

\noindent Selwyn, N., \& Gorard, S. (2016). Students' use of Wikipedia as an academic resource — Patterns of use and perceptions of usefulness. \textit{The Internet and Higher Education}, \textit{28}, 28-34. \vskip 0.1in

\noindent Solove, D. J. (2006). A Taxonomy of Privacy.\textit{ University of Pennsylvania Law Review}, \textit{154}(3), 477. \vskip 0.1in

\noindent Stein, L. (2013). Policy and Participation on Social Media: The Cases of YouTube, Facebook, and Wikipedia. \textit{Communication, Culture \& Critique}, \textit{6}(3), 353-371. \vskip 0.1in

\noindent Stevenson, A. (Ed.). (2010). \textit{Oxford Dictionary of English}. Oxford University Press, USA. \vskip 0.1in

\noindent Stop Online Piracy Act Bill 2011 (US). \vskip 0.1in

\noindent Sydow, M., Baraniak, K., \& Teisseyre, P. (2017). Diversity of editors and teams versus quality of cooperative work: experiments on wikipedia. \textit{Journal of Intelligent Information Systems}, \textit{48}(3), 601-632. \vskip 0.1in

\noindent Szoldra, P. (2016, September 16). This is everything Edward Snowden revealed in just one year of unprecedented top-secret leaks. \textit{Business Insider}. Retrieved from \url{https://www.businessinsider.com.au/snowden-leaks-timeline-2016-9} \vskip 0.1in

\noindent Tedjamulia, S. J., Dean, D. L., Olsen, D. R., \& Albrecht, C. C. (2005, January). Motivating Content Contributions to Online Communities: Toward a More Comprehensive Theory. In \textit{Proceedings of the 38th Hawaii International Conference on System Sciences} (pp. 193b-193b). IEEE. \vskip 0.1in

\noindent Tran, C., Champion, K., Forte, A., Hill, B. M., \& Greenstadt, R. (2020, May). Are anonymity-seekers just like everybody else? An analysis of contributions to Wikipedia from Tor. In \textit{2020 IEEE Symposium on Security and Privacy (SP)} (pp. 186-202). IEEE. \vskip 0.1in

\noindent Turnbull, S., \& Guthrie, J. (2019). Simplifying the Management of Complexity: As Achieved in Nature. \textit{Journal of Behavioural Economics and Social Systems}, \textit{1}(1), 51-73. \vskip 0.1in

\noindent Vrandečić, D. (2019). Collaborating on the sum of all knowledge across languages. \textit{Wikipedia @ 20}. \vskip 0.1in

\noindent Wales, J., \& Tretikov, L. (2015, March 10). Stop Spying on Wikipedia Users. \textit{The New York Times}. Retrieved from \url{https://www.nytimes.com/2015/03/10/opinion/stop-spying-on-wikipedia-users.html} \vskip 0.1in

\noindent Werbin, K. C. (2011). Spookipedia: intelligence, social media and biopolitics. \textit{Media, Culture \& Society}, \textit{33}(8), 1254-1265. \vskip 0.1in

\noindent \textit{Wikimedia Foundation, et al. v. National Security Agency, et al.} (2015). \vskip 0.1in

\noindent Wikipedia:Advice to users using Tor. (n.d.). In Wikipedia. Retrieved August 19, 2020, from \url{https://en.wikipedia.org/wiki/Wikipedia:Advice_to_users_using_Tor} \vskip 0.1in

\noindent Wikipedia:Manual of Style/Words to watch. (n.d.). In Wikipedia. Retrieved August 19, 2020, from \url{https://en.wikipedia.org/wiki/Wikipedia:Manual_of_Style/Words_to_watch} \vskip 0.1in

\noindent Wikipedia:SOPA initiative. (n.d.). In Wikipedia. Retrieved August 19, 2020, from \url{https://en.wikipedia.org/wiki/Wikipedia:SOPA_initiative} \vskip 0.1in

\noindent Wikipedia:Statistics. (n.d.). In Wikipedia. Retrieved August 19, 2020, from \url{https://en.wikipedia.org/wiki/Wikipedia:Statistics} \vskip 0.1in

\noindent Wikipedia:Wikipedia as a court source. (n.d.). In Wikipedia. Retrieved August 19, 2020, from \url{https://en.wikipedia.org/wiki/Wikipedia:Wikipedia_as_a_court_source} \vskip 0.1in

\noindent Wikipedia:Wikipedia in judicial opinions. (n.d.). In Wikipedia. Retrieved August 19, 2020, from \url{https://en.wikipedia.org/wiki/Wikipedia:Wikipedia_in_judicial_opinions} \vskip 0.1in

\noindent Xie, C. X., Johnson, I., \& Gomez, A. (2019, May). Detecting and Gauging Impact on Wikipedia Page Views. In \textit{Companion Proceedings of The 2019 World Wide Web Conference} (pp. 1254-1261). \vskip 0.1in

\end{document}